\documentclass[journal]{IEEEtran}
\usepackage{cite}
\ifCLASSINFOpdf
\else
\fi
\usepackage{amsmath}
\usepackage{graphicx}
\usepackage{amsmath}
\usepackage{bm}
\usepackage{amssymb}
\usepackage{booktabs}
\usepackage{multirow}
\usepackage{tabularx}
\usepackage{url} 
\begin{document}
\title{MLOps-Assisted Anomalous Reflector Metasurfaces Design Based on  Red Hat OpenShift AI}

\author{Wael Elshennawy}

\markboth{ }%
{Shell \MakeLowercase{\textit{et al.}}: Bare Demo of IEEEtran.cls for IEEE Journals}

\maketitle

\begin{abstract}
The integration of artificial intelligence as a design tool for metasurfaces, and the implementation of a deep-learning model pose a challenge in the development of an automated solution due to high resources requirements. The presented work introduces a network-layer solution to configure such environment for end user objectives, and for an underlying physical-layer technology. An architecture is developed to design an anomalous reflector by employing the Redhat Openshift AI (RHOAI) technology to support an automated machine learning operations (MLOps) framework in smart radio environments. This entails the design of lossless impenetrable metasurfaces characterized by a scalar surface impedance for an optimal anomalous reflection, achieved by optimizing the number of the Floquet modes through the utilization of a local power conservation constraint qualified as a fitness function. The metasurfaces design process is implemented by using a conditional generative adversarial network (cGAN). An extended cGAN with a surrogate model assists in a high-quality freeform metasurfaces design, where it introduces a swift simulation tool for the metasurfaces design process and analysis of the far-field model. The paper focuses on the challenges of building such a system, and potential abstraction layers. The training accuracy value of the proposed model demonstrates the feasibility and benefits of deploying in containerized environment of Red Hat Openshift in comparison with other deployments of ResNet-50 reported in literature.
\end{abstract}


\IEEEpeerreviewmaketitle

\section{Introduction}

{P}{reprogrammable and reconfigurable metasurfaces have contributed significantly to the evolution of established technologies. Nevertheless, they face a formidable challenge of quickly and automatically accommodating user-specific functionalities and site-specific requirements. Integration of intelligence into metasurfaces is essential for their adaptation to real-world environmental conditions and external stimuli \cite{liaskos2018realizing}, \cite{yang2025adaptively}, \cite{tsilipakos2020toward}. The primary goal aims at developing intelligent metasurfaces that can handle electromagnetic (EM) waves in a proficient way and autonomously perform intricate tasks. To realize this, intelligent metasurfaces must integrate three fundamental components: applications, intelligent algorithms, and tunable metasurfaces, which can be organized in a closed-loop framework \cite{liaskos2018using}.}

The overall EM response of metasurfaces cannot be accurately characterized solely by certain geometric principles due to the interdependence and coupling of constituent meta-atoms. Furthermore, in freeform metasurfaces, higher-order scattered waves such as second, third, or even more complex orders can be induced. Further complicating the issue, the freeform metasurfaces design process includes the characterization of an enormous amount of meta-atom designs with varying physical and geometric parameters, which demands numerous computational resources \cite{diaz2021macroscopic}. Consequently, there is a demand for streamlining the global design of metasurfaces to facilitate the implementation of intelligent algorithmic layers, which are essential advancement for the next generation of intelligent metasurfaces. In comparison with the design of a single unit cell or small-scale metamaterial, a major challenge arises from the increased dimensionality of the solution space.  Different solutions have been implemented using nonlocal response for perfect performance, such as  energy surface channeling in \cite{diaz2017generalized}, suppression of parasitic scattering  in \cite{movahediqomi2025simultaneous}, and  deep-learning networks in conjunction with a method to incorporate auxiliary surface waves in \cite{niu2023deep}.  

A field of research endeavors seeks to integrate deep learning (DL) to enhance the design performance of metasurfaces \cite{dai2022slmgan}, \cite{ma2019probabilistic}, \cite{niu2023deep}, which are intrinsically governed by physical laws. The utilization of neural networks to process inputs related to high-dimensional metasurfaces requires stringent demands on data generation, curation, and modeling of networks based on physical principles.  cGAN is promising for low-cost meatsurfaces design with complex frequency and polarization-dependent scattering responses \cite{10628337}. One of the most exciting features of cGAN is its ability to discover new geometries that have not been previously presented in the literature \cite{hodge2021deep}.   This suggests that the DL model implicitly learns the physical relationships of Maxwell's equations rather than simply interpolating from past designs. Thus, the trained cGAN model can generate on-demand meta-atom designs with given EM targets. The cGAN also employs a gradient penalty approach that is highly stable and easily converges. 

The field of DL fueled on-demand metasurfaces design has witnessed significant advancements, as it addresses the limitations of traditional numerical simulations and physics-based approaches in terms of inefficiency, time consumption, and reliance on experimental knowledge \cite{10172261}. However, rapid advancement in the development of intelligent metasurfaces is accompanied by certain challenges \cite{hodge2021deep}. Several critical questions arise, such as the initial collection of substantial amounts of labeled data, the frequent fine-tuning of network parameters, and the inefficacy in managing objectives with a higher degree of freedom. Moreover, the processes of sample collection and neural network training are intrinsically restricted to predefined individual metasurfaces and often falter when applied to larger-scale problems \cite{jia2023knowledge}, \cite{xu2025deep}.   

The ultimate goal is to automate the deployment of intelligent metasurfaces for applications featuring complex, dynamic, and unpredictable environments. The use of tunable standalone metasurfaces is insufficient for this purpose; hence, the metasurfaces must be integrated with a range of supplementary functionalities to develop a comprehensive perception-decision-reaction system. The emphasis is placed on the implementation of on-site learning and the exchange of knowledge across various settings. Red Hat OpenShift AI (RHOAI) can support this deployment and help to build a MLOps platform to create and deliver GenAI and predictive model by providing supported AI tooling on top of OpenShift for the automation of the metasurfaces operation lifecycle.  This would enable data scientists and developers to efficiently collaborate and create solid, scalable, and reproducible data science projects in response to this demand.

The paper presents MLOps assisted inverse design of the metasurfaces from desired wavefronts to obtain rapid designs. Lossless metasurfaces characterized by a scalar surface impedance are designed for perfect anomalous reflection based on field optimization. A set of co-polarized surface waves (SWs) introduced, and the total tangential fields are optimized such that the net power density normal to the surface is minimized everywhere toward obtaining a purely reactive surface impedance.  The DL model consists of the generator model and discriminator model of cGAN and additionally a surrogate model as an EM response prediction tool to find multiple structures with similar EM responses.  

MLOps employs a framework of practices and tools for managing the entire lifecycle of machine learning model used in metasurfaces design, enabling automation, reproducibility, and efficient deployment of this complex design process. The remaining parts of the paper are organized as follows. In Section II, the problem formulation is presented and the motivation for the proposed design framework is discussed, followed by the proposed procedure detailed the design framework including the inverse cGAN model and the HyperSurface-based programmable wireless environment model incorporating the  network layer of the software-defined networking (SDN) infrastructure is presented in Section III. The automation of cGAN model employing MLOps is discussed in Section IV. In Section V, the performance metrics of the ResNet 50 model implemented by RHOAI is presented for comparison with non-containerized environments. Section VI concludes the paper.

\section{Problem Formulation}

\begin{figure}
    \centering
    \includegraphics[width=1.0\linewidth]{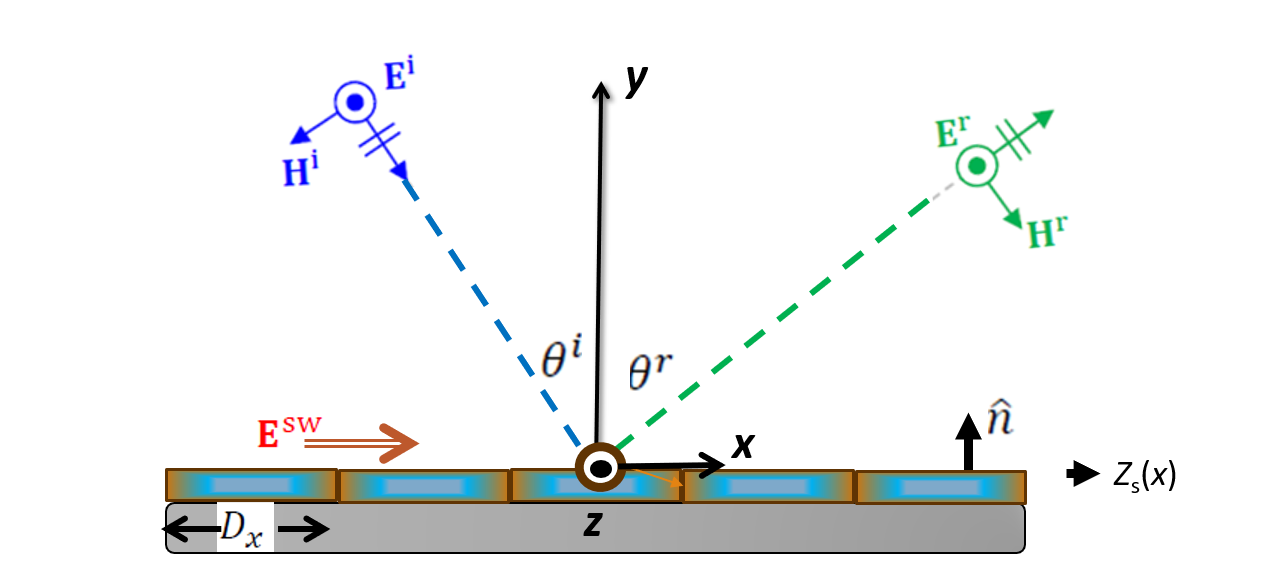}
    \caption{Plane-wave reflection in the TE$_y$ by an anomalous reflector metasurfaces in the $\textit{xz}$-plane.  Surface waves $\mathbf{E}^{SW}$ are co-polarized with the plane waves, adapted from \cite{kwon2018lossless}.}
    \label{Fig1} 
\end{figure}

In the context an anomalous reflector governed by the generalized reflection law, it has been demonstrated that the power efficiency directed towards the anomalous angle significantly decreases as the deflection angle increases \cite{7746401}. When the sum of the incident and reflected waves is defined as the total field, the resulting reflective surface displays alternating active and lossy spatial regions, yet it maintains average power neutrality. For $\Omega$ -bianisotropic metasurfaces, the incorporation of auxiliary evanescent waves in addition to propagating waves (PWs) and the enforcement of a local power conservation constraint yielded passive and lossless surface specifications conducive to a perfect anomalous reflection  \cite{PhysRevB.96.085438}. A polarization-preserving anomalous reflector of a plane wave is depicted in \ref{Fig1}, where an incident plane wave with fields ($\mathbf{E}^i,\mathbf{H}^i$) illuminates a reflector surface in the $xz$-plane with an incident angle $\theta^i$. Assume that all incident power is fully reflected in an anomalous reflection angle $\theta^r$ with plane wave fields ($\mathbf{E}^r,\mathbf{H}^r$).  

A set of co-polarized surface waves (SWs) is introduced, and the total tangential fields ($E_{tz},H_{tz}$) are then optimized to minimize the net power density  normal to the surface, with the goal of achieving a purely reactive surface impedance. The number of surface waves is optimized so that. the scalar surface impedance $Z_s (x)$characterizing the reflecting surface approaches a purely reactive profile. The no loss-condition is equivalent to the normal component of the Poynting vector satisfying $S_y(x)$ as follows in \cite{diaz2017generalized}.

\begin{align}\label{PF2}
S_y(x) = \frac{1}{2} \mathfrak{Re}\{E_{tz}H^{*}_{tx}\} = 0.    
\end{align}

This anomalous reflecting metasurfaces allow to further understand the efficiency and the energy distribution of the parasitic reflections for the design condition. In contrast, conventional numerical methods for freeform metasurfaces are time-consuming and require specialized expertise \cite{diaz2017generalized} . Recent studies demonstrate that DL has great potential to accelerate and refine metasurfaces design \cite{naseri2021combined}, \cite{niu2023deep}, \cite{zhang2019machine}. The importance of having ML strategies creates reusable patterns for production ready inference service and allows further automating the full ML model life cycle using cloud-native tooling and architectures. 

\section{Proposed Procedure}

\begin{figure*}[ht]
    \centering
    \includegraphics[width=\textwidth,height=4cm]{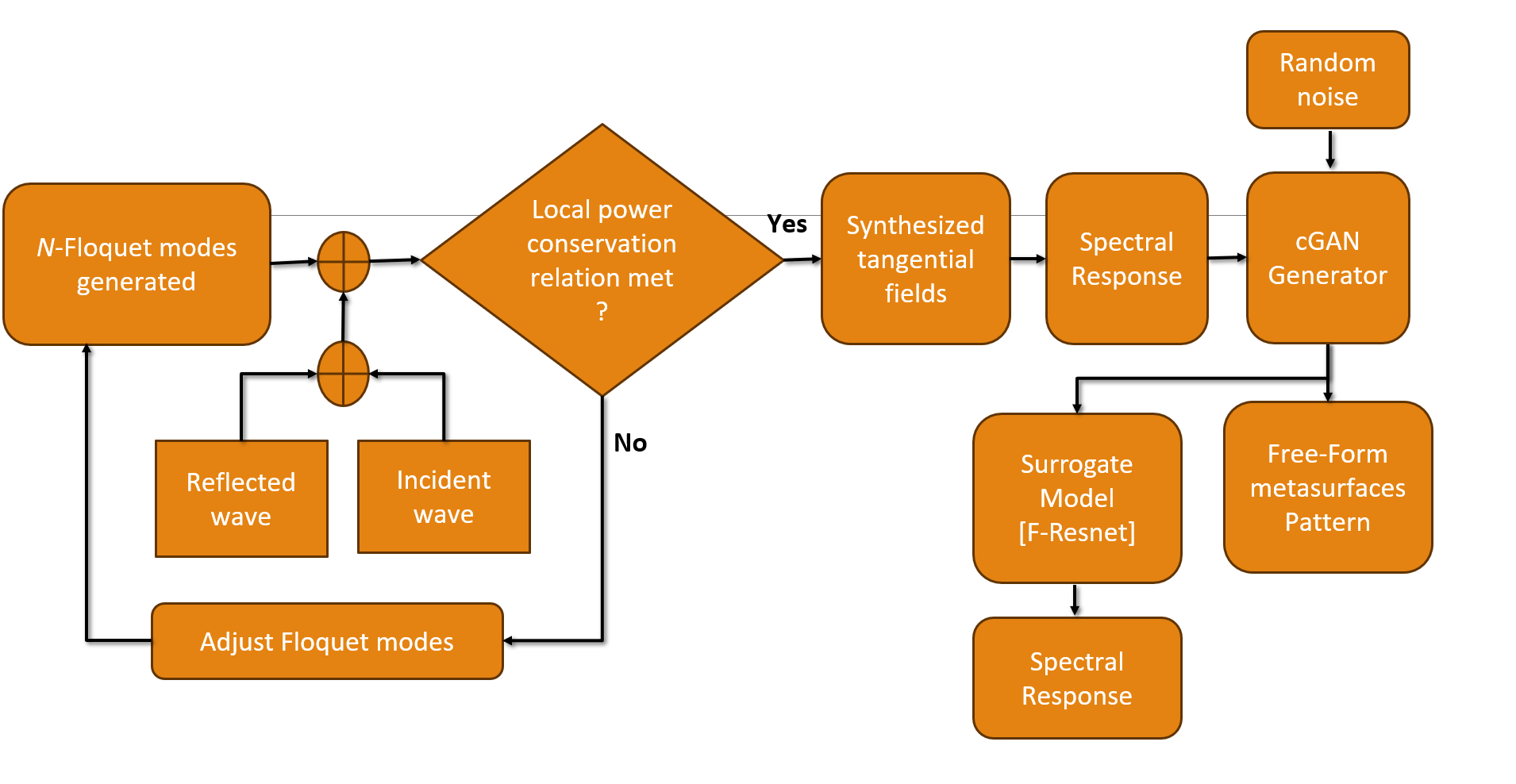}
    \caption{The training processes of the generator for inverse design of metasurfaces and the and the surrogate model.}
    \label{Fig2}
\end{figure*}

An approach is needed to increase computational efficiency and reduce the amount of training required for DL-based reconfigurable intelligent surfaces (RIS) design. This involves a DL model and an optimization architecture that utilizes a physics-based analytical model within the artificial neural network   (ANN) architecture \cite{niu2023deep}. Hybrid model is presented here, where the training dataset is fed by physics-based analytical model, that reduces the amount of required training dataset and enhances learning efficiency. This DL model-based analysis procedure aims to allow efficient synthesis in the form of iterative synthesis-analysis \cite{achouri2018design}. A DL model is developed in conjunction with the field optimization technique for the macroscopic design of an anomalous reflector. The optimization technique  employs the power redistribution fitness function to satisfy the requirement of having lossless and passive metasurfaces. 

The main input to the neural network model is the synthesized tangential fields, as illustrated in \ref{Fig2}. This network is used to calculate the freeform patterns of metasurfaces. According to Floquet theory, selecting the spatial periodicity of the structure in \ref{Fig1}, $D_{x}$, it allows to define the propagation directions of the reflected modes, Illuminating an infinite periodic metasurfaces by a plane wave at the incident angle $\theta_{i}$, the reflected field is defined as a superposition of plane waves propagating in different directions according to the relation in \cite{PhysRevB.96.085438}

\begin{align}\label{PF0}
sin\theta_{rn} = sin\theta_{i} + \frac{2\pi}{\kappa D_{x}} n,     
\end{align}

where $\kappa$ is the wavenumber and $\theta_{rn}$ are the reflection angles of the propagating harmonics of index $n = 0, \pm 1, \pm 2, ...$. Multiple evanescent modes are introduced and quantitatively controlled as illustrated in \ref{Fig2}, so the net power density normal to the surface is minimized everywhere in order to obtain a purely reactive surface impedance.  The total tangential fields are then optimized such that the local power conservation power relation is met. The synthesized tangential fields are then used to generate the required spectral responses. The cGAN model uses the desired spectral responses to guide and control the creation of specific freeform metasurfaces.  Adding contextual information helps the cGAN model to better structure the latent space, producing results that are more consistent, and aligned with expectations.

\subsection{cGAN-Driven Metasurfaces Inverse Design }

\begin{figure}
    \centering
    \includegraphics[width=1.0\linewidth]{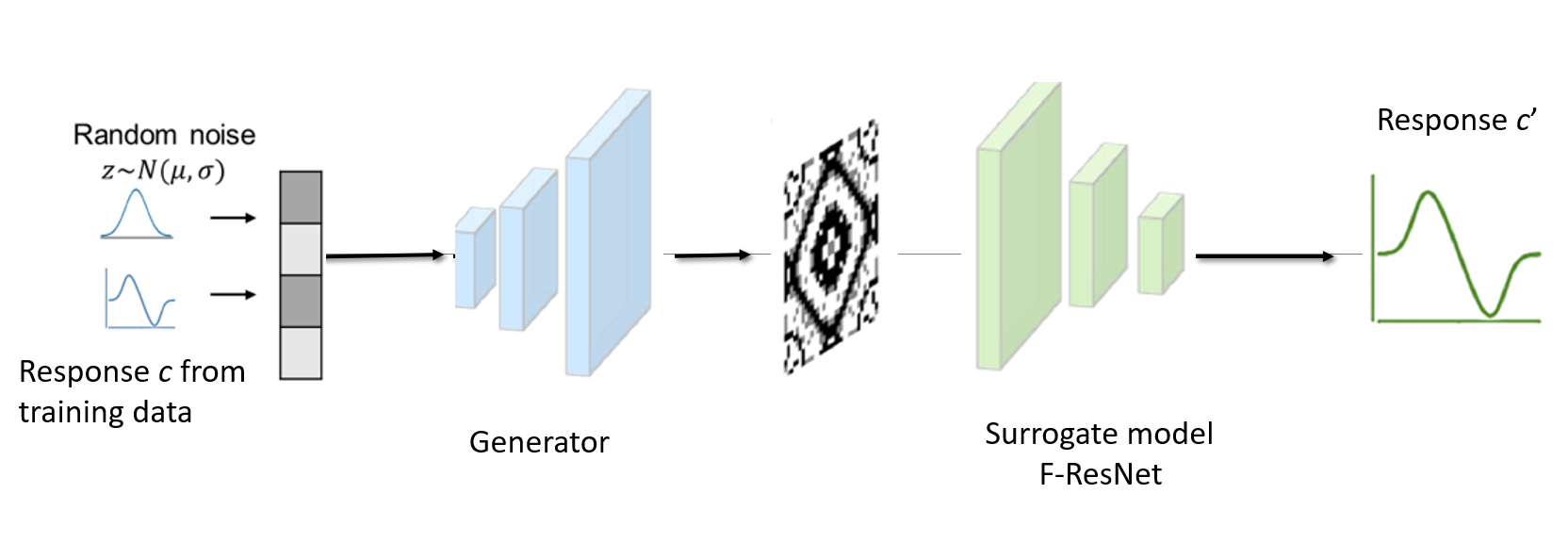}
    \caption{Modeling of cGAN, adapted from \cite{DAI2024106654}.}
    \label{Fig3}
\end{figure}

 The inverse design of metasurfaces geometry prediction  serves as one-to-many problem and can result in multiple patterns corresponding to the synthesized tangential fields. To address the one-to-many correlation between targets and designs, an extended cGAN model is guided by a pretrained surrogate model \cite{9382973}. This surrogate model leverages a cGAN functionality to automatically design freeform metasurfaces patterns that possess the desired spectral response.  It also serves as a prediction tool to analyze the metasurfaces pattern \cite{Jing:22}. The generator network has two components in the metasurfaces design framework: one is the surrogate model (Forward ResNet-50 \cite{pytorchResnet50x2014}) that predicts spectral responses from metasurfaces patterns, and the generator network that follows a specific sequence of layers and operations. This sequence involves generating patterns from a desired response as shown in Fig. \ref{Fig3}. During the training process, the desired response $c$ sampled from the training dataset is concatenated with random noise $z$ sampled from the Gaussian distribution $N(\mu, \sigma)$. This concatenated vector is fed into the generator and results in the desired pattern.  $c$ and $c'$ represent the desired response and the response generated by the surrogate model. 

 The learning process of a cGAN involves an adversarial game between two neural networks, a generator, and a discriminator. The architecture of the cGAN model is discussed in \cite{DAI2024106654}. During inference, only the generator is used, and its discriminator is removed. The discriminator serves to train the generator by evaluating the distance of the distributions between real patterns and synthetic patterns. At the end of the successful training process, the discriminator cannot distinguish batches generated by the generator and the training dataset \cite{hodge2021deep}. To generate an initial metasurfaces pattern, one feeds random noise following a Gaussian distribution and a target response into the generator. The Surrogate model is adopted to calculate the response of this pattern.  If the error between the calculated response $c'$ and the target response $c$ is below a predetermined threshold, the initial design pattern becomes the final design.  

Metasurfaces are the core technology for introducing programmable wireless environments (PWEs).  Using software capability to adjust metasurfaces functionality, where metasurfaces can support a wide range of EM interactions, denoted as functions.  A pertinent example; 6G-enabling systems that use RIS to turn radio propagation from an uncontrollable random phenomenon into customized network parameters \cite{BAE20241553}. To fulfill this requirement, an external software service calculates and deploys the optimal interaction types per metasurfaces to best fit the needs of communicating devices. This class of software-defined metasurfaces (SDMs) merge networked control elements with adaptive metasurfaces that will treat wireless propagation as an application \cite{liaskos2018new}, \cite{pitilakis2018software}.  The paper discusses here the function of reflection by implementing anomalous reflecting metasurfaces. In the next subsection, the integration of architecture and the associated complex underlying physics of SDMs are presented. Then, software abstractions are proposed that enable the interaction of the surfaces in a physics-agnostic manner.

\subsection{Software-Defined Metasurfaces and Programmable Wireless Environment}

\begin{figure*}
    \centering
    \includegraphics[width=\linewidth]{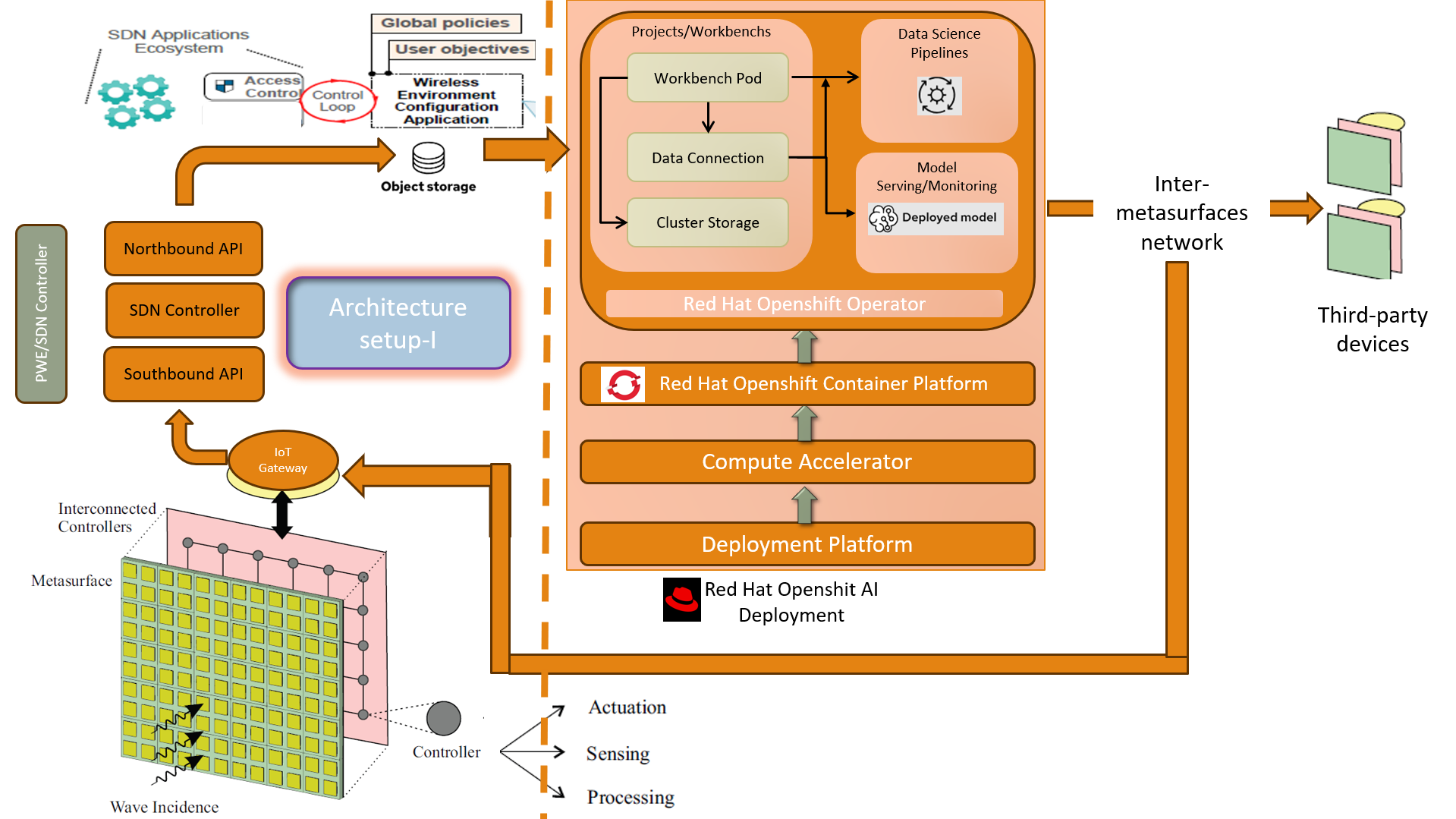}
    \caption{MLOps workflow for automating the design of metasurfaces in programmable wireless environments by integrating RIS/HyperSurfaces,  
SDN paradigm, and RHOAI. amended from \cite{1011453192336} } 
    \label{Fig4}
\end{figure*}

The PWEs are created by coating planar objects, such as walls and ceilings in an indoor environment, with tiles, i.e. rectangular panels of HyperSurfaces, with internetworking capabilities \cite{akyildiz20206g}. The latter allows a central server to connect to any tile, get its state and set its EM function in an automated manner. The paper discusses the physical layer and proposes an approach to build a MLOps platform at the multi-layered architecture of the PWEs. In other words, presenting an automated solution that incorporates user input with communication objectives within the PWEs as depicted in\ref{Fig4}, that seamlessly deploys, monitors and scales ML Models for real-world applications. Enhancing the Physical (PHY) layer with SDMs, a class of networked metasurfaces that comes with an application programming interface (API) and an EM compiler. The API allows for getting the HyperSurface state and setting its EM function, while abstracting the underlying physics \cite{8778748}. The EM compiler translates API callbacks to the corresponding active element states. The ability to change the desired output functionality is enabled by the adaptation of the metasurfaces to different wave characteristics. The local tunablility of the metasurfaces allows each constitutive unit cell of the metasurfaces be independently tuned. Additional functionalities are achieved by relying on the spatial modulation of the surface impedance of metasurfaces.  

Next to the EM behavior layer and the computing layer is the communication stack that consists of connection means between the actuating and sensing layers, the computing layer, and external tile devices.  This layer is simply implemented within the computing hardware, acting as a gateway to the other tiles, using any common protocol \cite{liaskos2022software}.  The gateway is usually connected to a software-defined networking (SDN) controller to receive EM manipulation commands via the EM API. The SDN controller discussed in the literature \cite{8466374} generally relies on the device localization system (DLS) to  know the floorplan schematics and the location of the metasurfaces in order to operate. The metasurfaces can play this role by finding graph paths employing the  ML model in \cite{elshennawy2012direction}, \cite{tsilipakos2020toward} The metasurfaces act as a very effective DLS out-of-the-box, which can accurately estimate the location of a user device, whilst performing other EM functions simultaneously. The PWEs also host a beacon, which is also wired to the SDN controller, but the discussion here is mainly limited to a single metasurfaces-tile scenario. Since the control and data planes are isolated in SDN, this makes network intelligence logically centralized.  

 The PWEs/SDN controller supports the main APIs, i.e., Northbound and Southbound. The Southbound API is responsible for controlling heterogeneous PWEs and SDN devices, including RISs, and other managed network devices. It also provides a uniform API to the controller components, where it employs designated adapters to the SDMs interface to the RISs \cite{liaskos2022software}. The Northbound API is responsible for providing “wireless channel as a service” capabilities to network services and management functions. Northbound APIs are used to expose the functionality of the SDN controller to higher-level network management applications and services, as shown in \textbackslash{}ref\{Fig4\}. These APIs allow external applications to interact with the SDN controller, request network services, and retrieve information. Northbound APIs enable programmability and automation of network management tasks and facilitate integration with orchestration systems, cloud platforms, and other management tools. Automation uses Northbound APIs for orchestration and Southbound APIs to programmatically configure, manage, and optimize traffic in real-time.

The SDN controller supports a standard interface between controller applications and forwarding tables within switches. Thus, it introduces a natural environment for network virtualization. The applications are normally residing in private or public cloud, which allows the itinerant users to access RIS from everywhere. However, this architecture setup-I is not prepared for high domain employment scenarios of cloud computing (left to the dashed line in \ref{Fig4}). It has no ability to manage demands, e.g. automation,  etc. In a cloud paradigm, it has become extra complicated and costly to offer novel end-to-end services and applications.  To cope with simultaneous requirements for manageability and automation in the cloud computing paradigm, the provision of novel services such as automation and dependable application delivery in a dynamic IT paradigm can be further realized by integrating the technology of SDN and RHOAI platform.

\section{MLOps for Automating Metasurfaces Design}

The end vision is to develop an Internet-of-Metasurfaces (IoM) \cite{10261137}, \cite{https://doi.org/10.1049/cmu2.12364}, with automated control over the electromagnetic behavior of materials.  The SDN controller supports high-level APIs that provide an abstraction of the network itself, such that the developer will not be concerned about dealing with network equipments separately but with the entire network.  Thus, by using the SDN controller, the network administrator can introduce a novel function by organizing software that can easily affect the logical map of the network slicing. Thus, this allows us to employ a framework of practices and tools to manage the entire lifecycle of the DL model deployed in the RHOAI platform. This permits us to employ it in metasurfaces design, enabling automation, reproducibility, and efficient deployment.

The architecture of RHOAI is illustrated in \ref{Fig4} to build out MLOps platform and deliver cGAN perdictive model supported by AI toolkit on top of Openshift \cite{brigoli2024mlops}. The model serving uses the data connection component of project/workbenches to download the cGAN model from S3-compatiable storage. After downloading the model, it exposes it for use via the REST API. The automation of ML model is carried out by the data science pipeline, where the workflow includes gathering real-data from data storage fed by real-data collected by SDN controller via RISs, data cleansing, training the cGAN model, evaluating the model, and finally stores saving model into S3-bucket storage for future retrieval.  The scalability of the cGAN model is managed by the workbench which autoscales the number of pods to accommodate the change of workload.  The availability to retain the data is ensured by the cluster storage using persistent volume claim (PVC) for data recovery in case of workbench restart. Finally, the dashboard and model monitoring can keep tabs on the performance of data science workloads and fix any potential problems. The RHOAI AI streamlines the entire MLOps lifecycle of the cGAN model by combining data science tools with DevOps principles and Kubernetes orchestration. 

\section{Performance metrics}

Comparing the computational efficiency of RHOAI with other non-containerized environments would require benchmarking on a specific dataset and hardware configuration. The dataset used in the performance comparison is an ImageNet. It is widely used for training and benchmarking in the field of computer vision, particularly for image classification and object detection tasks \cite{zhu2021phase} .   The primary metric for RHOAI is "performance delta" compared to non-containerized environments, i.e. platform overhead, optimization, etc. The computational efficiency of the cGAN-driven inverse metasurfaces design built by ResNet-50 employs feature extraction combined with a surrogate model for the EM response prediction tool varies merely based on several factors, including model architecture, hardware, and implementation details. In addition,  complexity is influenced by the size of the dataset, the number of training epochs, and the batch size. The training results of ResNet-50 on ImageNet implemented using OpenShift 4.13+ (the foundation for OpenShift AI) are within 0.4\% of the bare-metal results, which is near-native performance \cite{Feddema_2023}. 

Jia et al. achieved ResNet-50 model training in 6.6 minutes by using 2,048 GPUs \cite{jia2018highly}.  Ying et al. achieved 1.8 minutes using 1,024 TPU v3 processors \cite{ying2018image} .  The training time of ResNet-50 with 55,296 mini-batch size is 2.0 minutes is reported by Mikami et al. \cite{mikami2018massively} using Tesla V100 processors.  A minor reduction in training time is achieved by increasing the mini-batch size is reported by \cite{yamazaki2019yet}. For a single-node training for standard OpenShift AI has a range from 75.08 \% to 76.15 \% for Top-1 accuracy. It is clear that the training time and batch-size are the smallest with respect to the other models deployments. Recently, for advanced optimization using techniques like knowledge distillation or specialized training schedules (e.g. MEAL V2), vanilla ResNet-50 has been boosted to reach more than 80\% Top-1 accuracy. Table \ref{tab:placeholder} summarizes the performance of each model in key performance metrics: accuracy and training time. 
\begin{table*}[t]
    \centering
    \begin{tabular}{|c|c|c|c|c|c|}\hline
         Author&  Batch Size&  Processor&  DL Libaray&  Time (mins)& Accuracy (\%)\\\hline
         Jia et al. [40] &  65,536&  Tesla P40 × 2,048
&  TensorFlow&  6.6& 75.8
\\\hline
         Ying et al. [41] &  65,536&  TPU v3 × 1,024
&  TensorFlow&  1.8& 75.2
\\\hline
         Mikami et al. [42] &  55,296&  Tesla V100 × 3,456
&  NNL&  2.0& 75.29
\\\hline
        Yamazaki et al. [43] & 81,920& Tesla V100 × 2,048
& MXNet& 1.2&75.08
 \\\hline
         RHOAI-Standard Single-Node training
&  1,024&  4x NVIDIA A10G &  PyTorch \& TensorFlow
&  < 1& 75.08-76.15\\\hline
 
    \end{tabular}
    \caption{Training time and Top-1 validation accuracy with ResNet-50 on Imagnet.}
    \label{tab:placeholder}
\end{table*}

The RHOAI also permits to monitor the training and testing errors in real-time through the following OpenShift AI components. Model Monitoring: Integrated dashboards track drift detection and accuracy loss for models served online, comparing live inference data against the original training distribution. Pipelines: Data scientists use the Runs tab in OpenShift AI Pipelines to visualize training/validation error curves (e.g., MSE or MAE) generated during the training lifecycle. Resiliency.  

\section{Conclusion}

The paper presents a multi-layered architecture for intelligent metasurfaces in PWEs, comprising the algorithm layer, tunable metasurfaces layer, and application layer, and explores its implementation by utilizing RHOAI platform. RHOAI allows to automate and simplify the iterative process of integrating machine learning model into software development processes, production rollout, monitoring, retraining, and redeployment for continued prediction accuracy. The discussion entails anomalous reflector, cGAN model, underscoring its potential in the forward prediction and inverse design of freeform metasurfaces. The presented automated solution benefits form the Cloud-native edge computing which combines the speed of edge processing with cloud-native agility for efficient, scalable, and secure distributed applications. Furthermore, the paper identifies current challenges and delineates future perspectives to facilitate the advent of the era of intelligent metasurfaces. Because the suggested implementation simplifies the MLOps of the reconfigurable metasurfaces in a real environment, the realistic application of this model requires further consideration. The convergence of these technologies microservices and metasurfaces is one of the active research topics in the framework of 6G and Open RAN (O-RAN).  The paper focus on the importance of having ML strategies, creating reusable patterns for the production ready inference service, and automating the lifecycle of ML model using cloud-native tooling and architectures in the realm of intelligent metasurfaces. 

\bibliographystyle{IEEEbib}
\bibliography{references} 
\end{document}